\documentclass[useAMS,usenatbib,usegraphicx]{mn2e}
\usepackage{deluxetable}
\usepackage{journals}

\title[Size constraints on the jets of Sgr A* ]{How to hide large
  scale outflows: size constraints on the jets of Sgr A*}
\author[Markoff, Bower \& Falcke]{Sera Markoff$^{1}$\thanks{E-mail:
sera@science.uva.nl; gbower@astro.berkeley.edu; falcke@astron.nl}, Geoffrey C. Bower$^{2}$  and Heino Falcke$^{3}$\\
$^{1}$Astronomical Institute ``Anton Pannekoek'', University of
  Amsterdam, Kruislaan 403, 1098SJ Amsterdam, the Netherlands\\
$^{2}$601 Campbell Hall, Astronomy Department \& Radio Astronomy Lab,
  Berkeley, CA 94720, USA\\
$^{3}$ Department of  Astrophysics, Radboud University, Postbus 9010, 6500GL Nijgmegen;
  ASTRON, Postbus 2, 7990AA Dwingeloo, the Netherlands}
\begin{document}

\date{}

\pagerange{\pageref{firstpage}--\pageref{lastpage}} \pubyear{2002}

\maketitle

\label{firstpage}

\begin{abstract}
Despite significant strides made towards understanding accretion,
outflow, and emission processes in the Galactic Center supermassive
black hole Sagittarius A*, the presence of jets has neither been
rejected nor proven.  We investigate here whether the combined
spectral and morphological properties of the source at radio through
near infrared wavelengths are consistent with the predictions for
inhomogeneous jets.  In particular, we construct images of jets at a
wavelength of 7mm based on models that are consistent with the
spectrum of Sgr A*.  We then compare these models through closure
quantities with data obtained from the Very Long Baseline Array at
7mm.  We find that the best-fit jet models give comparable or better
fits than best-fit Gaussian models for the intrinsic source found in
previous analyses.  The best fitting jet models are bipolar, are highly
inclined to the line of sight ($\theta \ga 75^\circ$), may favor a
position angle on the sky of $105^\circ$, and have compact bases with
sizes of a few gravitational radii.
\end{abstract}


\section{Introduction}

Sgr A* is the compact radio source in our Galactic center, originally
discovered over 30 years ago by \citet{BalickBrown1974}.  Years of
dedicated observations of stellar orbits
\citep[e.g.,][]{Ghezetal2000,Schoedeletal2003} and precise,
high-resolution radio astrometry \citep{BackerSramek1999,Reidetal2003}
have led to Sgr A* being unambiguously associated with the central
supermassive black hole.  Most recently, the first measurements of the
instrinsic size of Sgr A* have been achieved
\cite{Boweretal2004,Shenetal2005,Boweretal2006}, giving information
about spatial structures extremely close to the black hole.  

For many years Sgr A* was only known to emit in the radio bands, with
a flat/inverted spectrum fairly typical of the compact cores of other
nearby low-luminosity galaxies \citep[e.g.][]{Ho1999,Nagaretal2002}.
However, the absence of infrared and higher energy emission was
puzzling given that at least some nonthermal accretion activity would
be expected for a source that is powered by weak accretion.  The first
positive identification of Sgr A* in the X-ray band with the {\em
Chandra Observatory} did not immediately settle the issue
\citep{Baganoffetal2001,Baganoffetal2003}.  The dominant quiescent
emission turned out to be extended and nonvariable, and thus likely
associated with hot gas within the Bondi capture radius of the black
hole \citep{Quataert2002}.  In contrast, the approximately daily
flares of nonthermal X-ray emission discovered later seem to originate
within tens of $r_g=GM/c^2$ from the black hole itself. Since this
discovery, Sgr A* has also been identified in the near infrared (NIR),
where it shows correlated variability with the X-ray band on similar
timescales \citep{Genzeletal2003,Ghezetal2004,Eckartetal2004}.  While
this suggests a low-level of active galactic nuclei (AGN)-like
behavior, the luminosity of Sgr A* ($\sim 10^{-9}L_{\rm Edd}$) is weak
enough to raise questions about comparisons with more luminous
accreting black holes.

Several models have been developed over the years to explain the
broadband emission of Sgr A*, ranging from Bondi-Hoyle infall
\citep{Melia1992}, to various radiatively-inefficient accretion flows
(RIAFs; \citealt{Narayanetal1998,YuanQuataertNarayan2003}), to jets
\citep{FalckeBiermann1995,FalckeMarkoff2000,Markoffetal2001}, and combinations thereof
\citep{YuanMarkoffFalcke2002}.  The persistence of such a wide range of
models can be attributed to some extent to the lack of constraints on
the nonthermal part of the X-ray spectrum.  Fitting the most compact
``submm bump'' region of the spectrum results in fairly similar
internal parameters for all current models, and this ``theoretical
degeneracy'' cannot easily be broken without better morphological
information from Very Large Baseline Interferometry (VLBI).
Unfortunately with current sensitivity and resolution limits, most
structure in the source is washed out by a strong scattering medium in
the central Galactic regions \citep[see, e.g.][]{Boweretal2006}.

Recently, however, several new observational techniques have been
developed which may help discern between various models. For instance,
the stringent limits placed on the accretion rate
($\dot{M}\sim10^{-9}-10^{-7}M_\odot$/yr) by measurements of linear
polarization
\citep{Aitkenetal2000,Boweretal2003,Boweretal2005,Marroneetal2006,Macquartetal2006}
have ruled out classical versions of the Bondi-Hoyle and
Advection-Dominated Accretion Flow (ADAF) models.  Similarly, better
determinations of the frequency-dependence of the electron scattering
law in the Galactic center (GC)
\citep{Boweretal2004,Shenetal2005,Boweretal2006} have resulted in new
constraints on models via their size-versus-frequency predictions.
While the different groups have found the index of the
size-versus-frequency relation to range from $\sim 1-1.6$, clearly any
successful model must be stratified (optically thick and thus having a
photosphere whose observable size varies with frequency) to achieve
this.  The determination of the scattering law to a high degree of
accuracy has allowed, for the first time, a dependable measurement of
the intrinsic size of Sgr A* along one axis as a function of
frequency.  This breakthrough, along with the expectation of
eventually determining the size in the other axis, means we are
finally at a key point where differences between models can be
empirically tested.  

In this paper, we use both the spectral data in combination with the
new VLBI measurements of the source photosphere at 43 GHz (from
\citealt{Boweretal2004} plus one new observation, see below) in order
to place new constraints on jet models.  In Section~\ref{sec2}, we
expand on the motivations for this project, in Section~\ref{sec3} we
introduce the model, in Section~\ref{sec4} we explain the methodology
and summarize our results in Section~\ref{sec5}, and discuss our
conclusions in Section~\ref{sec6}.

\section{The Evidence for Jets in Sgr A*}\label{sec2}

Because no jet in Sgr A* has yet been directly imaged, it is important to first
discuss the evidence in favor of jets in Sgr A*.  The lack of a
resolved core/jet structure is not surprising given the low
luminosity of Sgr A*, which suggests a small angular size for the jet,
 and the scattering screen in our line of sight towards
Sgr A*, which obscures small structures.  
Previous modeling of the structure of Sgr A* has succeeded in separating the intrinsic
and scatter-broadened images of Sgr A* via a Gaussian parameterization
of the intrinsic size.  
A primary goal of this paper is to go beyond this simple 
parameterization.

In fact, there are several strong arguments for jets in Sgr A*.  On a
purely theoretical level, some form of jet production seems to go
hand-in-hand with accretion around black holes, both at the galactic
as well as stellar scales.  In stellar black holes accreting from a
binary companion, or X-ray binaries (XRBs), jet production is observed
to be cyclic over outburst cycles.  The strongest (relative to the
system energetics) and steadiest jets occur during the low-luminosity
state, called the Low/Hard State, while during the highest
luminosity state, the jets appear quenched \citep{Fenderetal1999b}.
The low-luminosity jets are compact and self-absorbed with a
flat/inverted spectrum, and correlated radio/X-ray variability has
demonstrated that the jets increasingly dominate the power output as
the luminosity decreases \citep{FenderGalloJonker2003}.  The weakest
accreting black hole we can study with reasonable statistics besides
Sgr A* is the XRB A0620-00, in which radio emission has recently been
detected \citep{Galloetal2006}.  At an X-ray luminosity of $\la
5\times10^{-9}$ $L_{\rm Edd}$, very close to that of Sgr A*, efficient
jets are still produced in this black hole, with characteristics
matching those at higher powers.  If general relativity's basic prediction
of scaling black hole physics holds, this is a strong argument for jet
production in Sgr A*.
 
The radio spectrum, radio variability, and
high-frequency linear polarization are all similar to other nearby
low-luminosity AGN (LLAGN; \citealt{Ho1999,FalckeBiermann1999,
BowerFalckeMellon2002,Nagaretal2002,NagarFalckeWilson2005}).
Most of the observed cores are accretion-powered, and have the
signature flat/inverted, self-absorbed radio spectrum associated with
compact jets \citep{BlandfordKoenigl1979}.  While the jets can only
generally be resolved in the brightest sources, when they are resolved
they dominate the unresolved core by at least a factor of a few.  The
results of these surveys strengthen the arguments for a jet in Sgr A*
based on its radio spectrum and polarization.

One source that is particularly interesting because of its many
parallels with Sgr A* is the nucleus of the nearby LLAGN M81.  M81* is
our nearest LLAGN besides Sgr A*, and resides in the same kind of
spiral galaxy as the Milky Way.  Its mass has been derived from line
spectroscopy (using {\em HST}; Devereux et al. 2003) to be
$7\times10^7$ $M_\odot$, only $\sim 30$ times the mass of Sgr A*.
M81* also possesses the typical compact flat/inverted core spectrum
(\citealt{Falcke1996}; Markoff et al., in prep.)  and, more
importantly, the same high levels of circular rather than linear
polarization in the centimeter radio band as Sgr A*
\citep{Brunthaleretal2001,BrunthalerBowerFalcke2006}.

The M81* jet is one-sided, very small (700-3600 AU depending on the
frequency, with a roughly $\sim 1/\nu$ dependence), and exhibits
occasional bends in its morphology \citep{BietenholzBartelRupen2000}.
Scaling the size by mass alone would argue for a $\sim 20-120$ AU jet
in Sgr A*, but the observed size should also scale with luminosity,
depending on the particulars of the jet model and frequency.  In fact,
the jet nature of M81* was difficult to establish due to the high
level of compactness.  Taking into account Sgr A*'s five orders of magnitude
lower power, as well as the scattering screen, it is not surprising
that no jet has yet been detected in our Galactic center.

Another argument in favor of jets comes from the recent detection of
short time delays of about 0.5-1 hr between 43 and 22 GHz for waves of
variability traveling from high to low frequencies
\citep{Yusef-Zadehetal2006b}.  This variability is fully consistent
with outflowing, adiabatically expanding blobs of plasma, as would be
expected for jets (in fact, the model the authors use to interpret
their results was developed in this context).

Finally, the recent size-versus-frequency scaling detections support
an optically thick, stratified model such as a self-absorbed jet.
While the predictions of the jet model presented in
\cite{FalckeMarkoff2000}, as well as that of RIAFs
\citep{YuanShenHuang2006}, are consistent with a $1/\nu$ scaling, and
thus with the results in \cite{Boweretal2004} and \cite{Shenetal2005}, they
disagree with the steeper index determined more recently by
\cite{Boweretal2006}.  If this latter result is indeed correct, it
suggests that the current versions of all models, jets included, need
to be modified to show a stronger dependence on observing frequency.
Because this issue is still under debate, however, in this paper we
are still using the original scaling relation.  

Although the circumstantial evidence is significant, there are
other complications which could argue against jets.  For instance,
XRBs in their steady-jet producing Low/Hard state display a
correlation between their radio and X-ray luminosities that holds over
at least seven orders of magnitude in luminosity
\citep{Corbeletal2003,GalloFenderPooley2003}.  Among other things,
this correlation can be used as a gauge for ``typical'' levels of
activity.  The recent radio detection of A0620-00 falls exactly on the
correlation, extending it to even lower luminosities and indicating
that the same mechanism is at work as in brighter sources where jets
can be imaged.  If the physics driving the correlation scales in a
predictable way with mass, it should apply to LLAGN as well, where the
mass enters mainly as a normalization factor for the same correlation slope.
This relationship between radio and X-ray luminosities and mass is
called ``the fundamental plane of black hole accretion'' and has been
explored in several recent papers
\citep{MerloniHeinzDiMatteo2003,FalckeKoerdingMarkoff2004,KoerdingFalckeCorbel2006,Merlonietal2006}.
When Sgr A* in quiescence is placed on this plane, it falls well below
the correlation in predicted X-rays, given its radio luminosity. One
could interpret this as the complete dominance of the jet over inflow
processes at the lowest of luminosities, but it could also mean that
the emission mechanisms themselves have undergone a transition to a
different mode of emission entirely.

In order to try to cast new light on these long-standing ambiguities
and place more stringent constraints on the possible presence of jets
in Sgr A*, we have developed a new method to combine spectral and
morphological data.  Our results will set the stage for future tests
with upcoming VLBI observations at millimeter and submillimeter
wavelengths, where the morphology is less affected by scattering and
resolution is comparable to a few $r_g$.

\section{Model}\label{sec3}

Like most models involving optically thick, collimated outflows, we
build on the initial work of \citet{BlandfordKoenigl1979}.  These
authors demonstrated the ``conspiracy'' of how a perfectly flat
spectrum ($\alpha\sim0,\;F_\nu\propto\nu^{-\alpha}$) can result from a
superposition of self-absorbed contributions along a conical,
idealized jet.  When more realistic physics such as bulk acceleration,
full particle distributions and cooling are included, compact jets show a
slight spectral inversion in the radio wavebands, with $\alpha\sim
0.0-0.2$.  The model used here is based on a model developed for Sgr
A* \citep{FalckeMarkoff2000}, which has been significantly modified to
extend to XRBs and LLAGN in general.  For a detailed description see
the appendix in \citet{MarkoffNowakWilms2005}; we provide only a
brief summary below.  

The model is based upon four assumptions: 1) the total power in the
jets scales with the total accretion power at the innermost part of
the accretion disk, $\dot{M}c^2$, 2) the jets are freely expanding and
only weakly accelerated via their own internal pressure gradients, 3) the
jets contain cold protons which carry most of the kinetic energy while
leptons dominate the radiation and 4) particles have the opportunity to
be accelerated into power-law tails.  In sources accreting at higher
levels this latter point would be more important, but as we will show
later, there is not much capacity in the Sgr A* spectrum for significant
particle acceleration.  

The base of the jets consist of a small nozzle of constant radius
where no bulk acceleration occurs.  The nozzle absorbs our
uncertainties about the exact nature of the relationship between the
accretion flow and the jets, and fixes the initial value of most
parameters.  Beyond the nozzle the jet expands laterally with its
initial proper sound speed for a relativistic electron/proton plasma,
$\gamma_{\rm s}\beta_{\rm s}c\sim0.4c$.  The plasma is weakly
accelerated by the resulting longitudinal pressure gradient force,
allowing an exact solution for the velocity profile via the Euler
equation \citep[see, e.g.,][]{Falcke1996}.  This results in a roughly
logarithmic dependence of velocity upon distance from the nozzle, $z$.
The velocity eventually saturates at large distances at Lorentz
factors of $\Gamma_{\rm j}\ga$2-3.  The size of the base of the jet,
$r_0$, is a free parameter (but expected to fall within several $r_g$)
and once fixed determines the radius as a function of distance along
the jet, $r(z)$.  There is no radial dependence in this model.

\begin{figure}
\centerline{\includegraphics[width=0.5\textwidth]{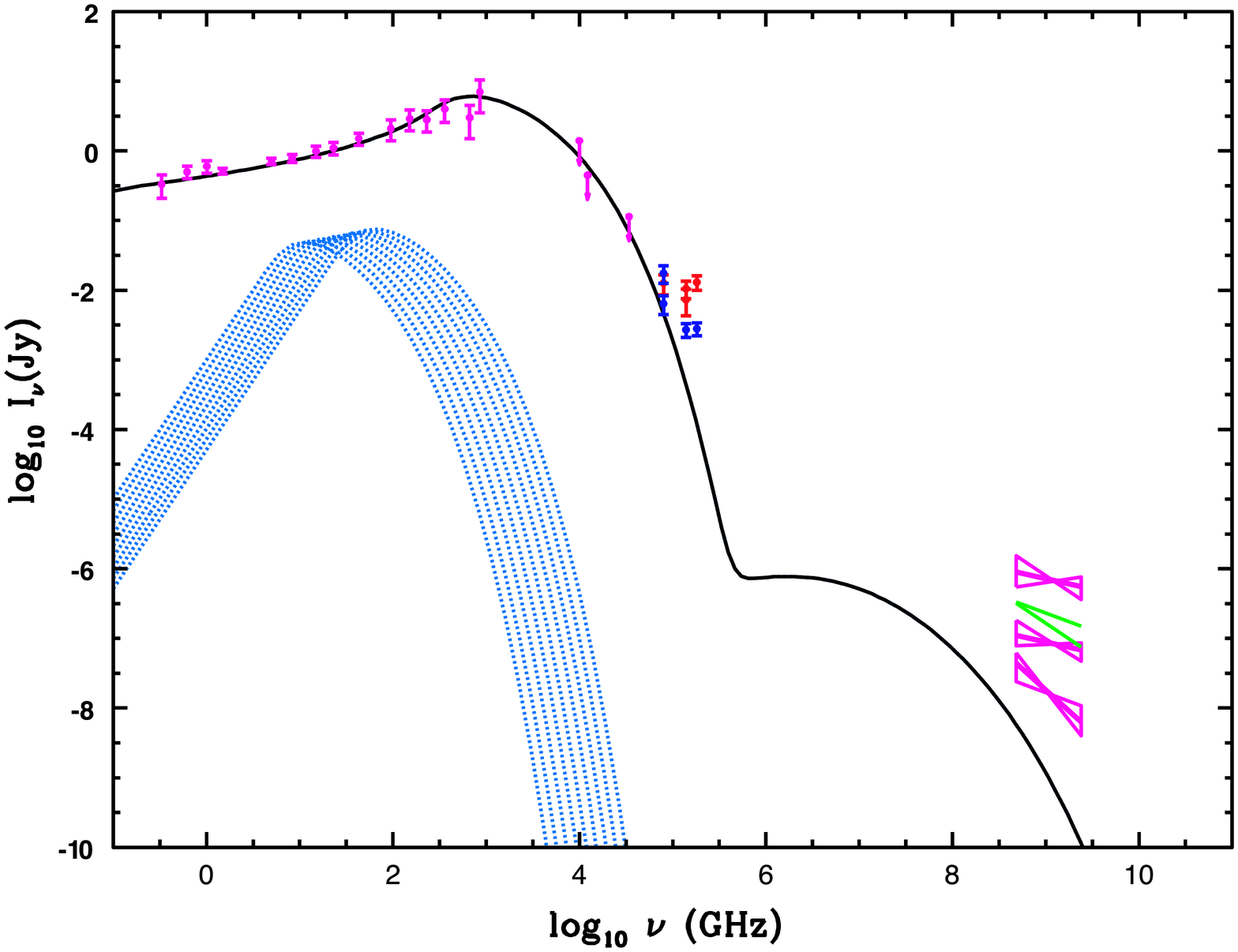}}
\caption{The broadband data set for Sgr A* used to constrain our
  models, taken from the average spectrum up until the submillimeter
  \citep{MeliaFalcke2001}, with additional low frequency
  points from \citet{Nordetal2004} and \citet{RoyRao2004} and infrared data from
  \citet{Genzeletal2003} and \citet{Ghezetal2004}.  The X-ray ``bow-ties''
  represent the quiescent (lowest), average daily {\em Chandra} flare
  (middle) and brightest {\em Chandra} (top) power-laws with errors
  indicated \citep{Baganoffetal2001,Baganoff2003}.  The ``V'' shape
  indicates the two {\em XMM-Newton} flares presented in
  \citet{Belangeretal2005}.  The solid curve shows a representative
  quiescent model with synchrotron and synchrotron self-Compton peaks.
  The dotted lines illustrate the contribution of the quasi-thermal
  particles from each increment along the jet, which superimpose to
  give the characteristic flat/inverted synchrotron spectrum. }
\label{fig:sgraquiet}
\end{figure}

The model is most sensitive to the fitted parameter $N_{\rm j}$, which
acts as a normalization.  It dictates the power initially divided
between the particles and magnetic field at the base of the jet, and
is expressed in terms of a fraction of the Eddington luminosity
$L_{\rm Edd}=1.25\times10^{38} M_{\rm bh,\odot}$ erg s$^{-1}$.  Once
$N_{\rm j}$ and $r_0$ are specified and conservation is assumed, the
macroscopic physical parameters along the jet are determined.  We
assume that the jet power is roughly shared between the internal and
external pressures.  The radiating particles enter the base of the jet
where the bulk velocities are lowest, with a quasi-thermal
distribution.  In higher power jets, a significant fraction of the
particles are accelerated into a power-law tail, however in Sgr A*
this seems to be less of an effect.  The particles in the jet
radiatively cool via adiabatic expansion, the synchrotron process, and
inverse Compton upscattering; however, adiabatic expansion is assumed
to dominate the observed effects of cooling.  Because Sgr A* has no
``standard thin accretion disk'' \citep[e.g.][]{ShakuraSunyaev1973},
nor even a fossil disk, which would be apparent in the infrared
\citep{FalckeMelia1997}, the photon field for inverse Compton
upscattering is entirely dominated by locally produced synchrotron
photons.  Fig.~\ref{fig:sgraquiet} shows an example of the resulting
broadband spectrum plotted against the data for Sgr A*.  

Besides those mentioned above, the other main fitted parameters are
the ratio of length of the nozzle to its radius $h_0$, the electron
temperature $T_e$, the inclination angle between the jet axis and
line of sight $\theta_i$ and the equipartition parameter between the
magnetic field and the radiating (lepton) particle energy densities,
$k$.  

\begin{figure}
\centerline{\includegraphics[width=0.5\textwidth]{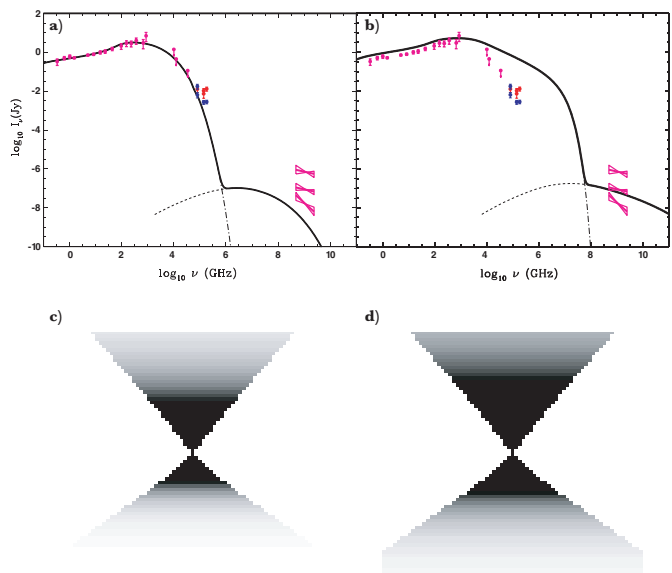}}
\caption{Illustration of the effect of particle acceleration on the
  observed jet profile.  Panel a) shows the quiescent model 3, which
  provides the best statistical description of the radio data.  Panel
  b) shows a model with the exact same parameters except for the
  addition of significant particle acceleration, where 75\% of the
  particles in the quasi-thermal distribution are accelerated into a
  power-law with energy index 2.2, and a cutoff 50 times higher than
  the minimum energy (which is fixed to the peak of the thermal
  distribution).  Panels c) and d) show the profile this model
  produces on the sky, prior to convolution with the scattering
  Gaussian.  The scale of the images is 1 mas.  The images are
  saturated to emphasize the brightest (darkest) parts of the jets.  A
  tail of accelerated particles serves to significantly lengthen the
  jets profile at a given frequency.}
\label{fig:model3comp}
\end{figure}

Aside from Sgr A*, this class of model has been successfully applied
to several LHS XRBs
(\citealt{MarkoffFalckeFender2001,Markoffetal2003,MarkoffNowakWilms2005,Migliarietal2007};
Gallo et al., in prep.) and other LLAGN (\citealt{Yuanetal2002}; Filho
et al., in prep., Markoff et al., in prep.).  As would be expected
from the existence of the fundamental plane, all significantly
sub-Eddington accreting black holes do seem to share some basic
underlying physics across the mass scale.  However, as mentioned
above, Sgr A* does not participate in the radio/X-ray correlation and
can only be reconciled into this picture if significant particle
acceleration is lacking.  This is a very interesting point, because
the appearance (or non-appearance) of a jet is strongly dependent on
its internal particle distributions.  A power-law tail of accelerated
particles results in more optically thin synchrotron emission over a
broader frequency range from each jet increment.  Thus when observing
at a single frequency, a larger range of increments are able to
contribute to the profile, resulting in a larger jet image, as we show
in Fig.~\ref{fig:model3comp}.

\section{Methodology}\label{sec4}

\subsection{Modeling the spectral data of Sgr A*}

In order to break the current degeneracy based on modeling the
spectral data alone, we are for the first time calculating the
predicted size and morphology of jet models which give a good
description of the broadband data.  After convolving these ``maps''
with the scattering screen, we then compare the results to closure
quantities from VLBI, which provide information about the structure.
The model predictions are analyzed in the same way as the
observations.  

The 41 models presented here were chosen to represent a range which
samples the full parameter space within the constraints of a
reasonably good ($\chi^2 < 1$) description of the radio through IR.
For quiescent models, they are further constrained to fall within a
factor of a few under the X-ray quiescent limits from {\em Chandra}.
Flaring models are those which can account for either the average
daily flaring flux or the highest detected with {\em Chandra} in the
X-ray band, via some form of heating, accelerating or increased power
compared to the quiescent state.  We initially conducted a very rough
search of a large region of parameter space for the first 20 models,
then focused on a smaller region to explore properties nearest the
best-fitting models, as well as for flares.  A systematic exploration
of the parameter space would be at least a 6 dimensional parameter
cube, which we deemed too computationally intensive for this initial
test study.

Because there is still significant uncertainty about what to consider
the ``quiescent'' versus ``flared'' IR flux amplitude and slope, we
did not include the IR in the $\chi^2$ calculation but rather required
the model to fall reasonably within range of the errors. The inability
to constrain the exact IR and X-ray quiescent flux accounts for almost
all of the allowed range in the fitted parameters for the quiescent
state, otherwise the parameters would be fairly tightly determined.
In this way the addition of morphological fitting can constrain the
quiescent contribution above the submm band.  We also compared our
models to the simultaneous radio through IR data set presented in
\citet{Anetal2005}, and found the level agreement between the two data
sets so good that we did not feel it provided an additional
constraint.  One of our secondary goals was to explore observable
differences in the photosphere during flaring states.

Table~\ref{tab:modpars} lists the models and their parameters, along
with relevant comments.  

\onecolumn
\begin{deluxetable}{rccccccccccc} 
\rotate
\setlength{\tabcolsep}{0.065in} 
\tabletypesize{\scriptsize} 
\tablewidth{0pt} 
\tablecaption{Jet model parameters \label{tab:modpars}} 
\tablehead{  \colhead{Model}
           & \colhead{Q/F}
           & \colhead{$N_{\rm j}$} 
           & \colhead{$r_{0}$}       
           & \colhead{$h_{0}/r_{0}$}      
           & \colhead{$\theta_i$}
           & \colhead{$T_{\rm e}$}          
           & \colhead{$k$}            
           & \colhead{$n_e=n_p$}  
           & \colhead{$n_j$}
           & \colhead{Other$^a$}
           & \colhead{$\chi^2$/DoF$^b$}\\
& & ($10^{-7}L_{\rm Edd}$) & ($r_g$) & & ($^\circ$) & ($10^{11}$ K) & &&&&}
\startdata 
1 & Q
            & $3.6$        
            & 5                         
            & 2.5                       
            & 51                        
            & $2.2$         
            & 10                        
            & y                         
            & 2                         
            &                           
            & 5.77/9                    
            \\ 
2 & Q
            & $6.2$        
            & 5                         
            & 1.5                       
            & 67                        
            & $1.9$         
            & 10                        
            & y                         
            & 2                         
            &                           
            & 8.48/9                    
            \\ 
3 & Q
            & $7.2$        
            & 5                         
            & 1.1                       
            & 75                        
            & $1.8$         
            & 15                        
            & y                         
            & 2                         
            &                           
            & 4.73/9                    
            \\ 
4 & Q
            & $5.8$        
            & 3                         
            & 1.3                       
            & 55                        
            & $1.8$         
            & 10                        
            & y                         
            & 2                         
            &                           
            & 7.05/9                    
            \\ 
5 & Q
            & $6.6$        
            & 3                         
            & 1.2                       
            & 52                        
            & $1.7$         
            & 10                        
            & y                         
            & 2                         
            & $\beta(z)$ stretched      
            & 4.98/9                    
            \\ 
6 & Q
            & $6.0$        
            & 3.5                         
            & 1.2                         
            & 45                        
            & $1.7$         
            & 10                        
            & y                         
            & 2                         
            & $\beta(z)$ stretched      
            & 4.74/9                    
            \\ 
7 & Q
            & $10$        
            & 5                         
            & 1.2                       
            & 65                        
            & $2.4$         
            & 2                        
            & y                         
            & 2                         
            &                           
            & 6.38/9                    
            \\ 
8 & Q
            & $7.2$        
            & 3.5                         
            & 1.7                       
            & 57                        
            & $1.6$         
            & 10                        
            & y                         
            & 1                         
            &                           
            & 4.88/9                    
            \\ 
9 & Q
            & $2.6$        
            & 3.5                         
            & 2.2                       
            & 50                        
            & $2.0$         
            & 10                        
            & n                         
            & 2                         
            &                           
            & 5.76/9                    
            \\ 
10  & Q
            & $1.8$        
            & 3.5                         
            & 1.9                       
            & 50                        
            & $2.1$         
            & 4                        
            & n                         
            & 2                         
            &                           
            & 5.27/9                    
            \\ 
11 & Q
            & $3.0$        
            & 3.5                         
            & 2.4                       
            & 50                        
            & $1.7$         
            & 10                        
            & n                         
            & 1                         
            &                           
            & 6.69/9                    
            \\ 
12 & Q
            & $6.6$        
            & 5                         
            & 1.1                       
            & 69                        
            & $2.0$         
            & 10                        
            & y                         
            & 2                         
            &                           
            & 5.54/9                    
            \\ 
13 & Q
            & $3.4$        
            & 8                         
            & 1.4                       
            & 63                        
            & $2.0$         
            & 30                        
            & y                         
            & 2                         
            &                           
            & 9.44/9                    
            \\ 
14 & Q
            & $2.4$        
            & 5                         
            & 1.8                       
            & 55                        
            & $2.1$         
            & 6                        
            & n                         
            & 2                         
            &                           
            & 5.67/9                    
            \\ 
15 & F
            & $16$        
            & 5                         
            & 1.3                       
            & 67                        
            & $7.0$         
            & 0.1                        
            & y                         
            & 2                         
            &      
            & 13.22/9                    
            \\ 
16 & F
            & $16$        
            & 5                         
            & 1.1                       
            & 75                        
            & $7.2$         
            & 0.1                        
            & y                         
            & 2                         
            & Av. {\em Chandra} flare, SSC     
            & 10.06/9                       
            \\ 
17 & F
            & $18$        
            & 5                         
            & 1.1                       
            & 75                        
            & $7.9$         
            & 0.05                        
            & y                         
            & 2                         
            & Av. {\em Chandra} flare, SSC             
            & 14.31/9                       
            \\ 
18 & Q 
            & $7.2$        
            & 5                         
            & 1.1                       
            & 85                        
            & $1.9$         
            & 15                        
            & y                         
            & 2                         
            &                           
            & 5.68/9                    
            \\ 
19 & F 
            & $18.4$        
            & 5                         
            & 1.1                       
            & 85                        
            & $6.5$         
            & 0.09                        
            & y                         
            & 2                         
            & Av. {\em Chandra} flare, SSC           
            & 10.15/9                       
            \\ 
20 & Q
            & $11$        
            & 3.5                         
            & 1.0                       
            & 85                        
            & $1.4$         
            & 15                        
            & y                         
            & 2                         
            &                           
            & 6.44/9                    
            \\ 
21 & Q
            & $7.6$        
            & 3.5                         
            & 1.0                       
            & 85                        
            & $1.4$         
            & 15                        
            & n                         
            & 2                         
            &                           
            & 5.69/9                    
            \\ 
22 & Q
            & $14$        
            & 5                         
            & 1.0                       
            & 85                        
            & $1.5$         
            & 15                        
            & y                         
            & 1                         
            &                           
            & 9.15/9                    
            \\ 
23 & Q
            & $10$        
            & 3                         
            & 1.0                       
            & 85                        
            & $1.5$         
            & 15                        
            & y                         
            & 2                         
            &                           
            & 8.69/9                    
            \\ 
24 & Q
            & $19$        
            & 2.5                         
            & 1.0                       
            & 85                        
            & $1.0$         
            & 15                        
            & y                         
            & 2                         
            &                           
            & 12.19/9                    
            \\ 
25 & Q
            & $20$        
            & 2                         
            & 1.0                       
            & 70                        
            & $0.8$         
            & 20                        
            & y                         
            & 2                         
            &                           
            & 12.66/9                    
            \\ 
26 & Q
            & $19$        
            & 2                         
            & 1.0                       
            & 80                        
            & $1.0$         
            & 15                        
            & y                         
            & 2                         
            &                           
            & 11.67/9                    
            \\ 
27 & Q
            & $13$        
            & 3                         
            & 0.6                       
            & 87                        
            & $1.3$         
            & 15                        
            & y                         
            & 2                         
            &                           
            & 5.46/9                    
            \\ 
28 & Q
            & $13$        
            & 3                         
            & 0.6                       
            & 87                        
            & $1.3$         
            & 15                        
            & y                         
            & 2                         
            & $z_{\rm acc}=50$,$p=3$,$u/f=7\times10^{-3}$,plf$=0.1$ 
            & 4.51/5                    
            \\ 
29 & Q
            & $13$        
            & 3                         
            & 0.6                       
            & 87                        
            & $1.3$         
            & 15                        
            & y                         
            & 2                         
            & $z_{\rm acc}=50$,$p=3$,$u/f=3\times10^{-4}$,plf$=0.1$                          
            & 4.52/5                    
            \\ 
30 & Q
            & $140$        
            & 5                         
            & 0.6                       
            & 85                        
            & $0.3$         
            & 15                        
            & y                         
            & 2                         
            & PL:$p=3.4$, $\gamma_{\rm e,max}=2\times10^3$                          
            & 7.10/7                    
            \\ 
31 & Q
            & $38$        
            & 3                         
            & 0.6                       
            & 85                        
            & $0.7$         
            & 15                        
            & y                         
            & 2                         
            & PL:$p=3.4$, $\gamma_{\rm e,max}=3\times10^3$                          
            & 8.23/7                    
            \\ 
32 & Q
            & $60$        
            & 3                         
            & 0.4                       
            & 85                        
            & $0.6$         
            & 15                        
            & y                         
            & 2                         
            & PL:$p=3.8$, $\gamma_{\rm e,max}=2.5\times10^3$                         
            & 3.83/7                    
            \\ 
33 & Q
            & $23$        
            & 2                         
            & 0.7                       
            & 85                        
            & $0.95$         
            & 15                        
            & y                         
            & 2                         
            &                           
            & 5.99/9                    
            \\ 
34 & F
            & $19$        
            & 2.5                         
            & 1.0                       
            & 85                        
            & $1.0$         
            & 15                        
            & y                         
            & 2                         
            & $z_{\rm acc}=10$, $p=1.7$, $u/f=0.014$, plf$=1\times10^{-4}$                          
            & 10.35/5                    
            \\ 
&&&&&&&&&& Biggest {\em Chandra} flare, synch. &\\
35 & F 
            & $11.5$        
            & 2.5                         
            & 0.95                       
            & 85                        
            & $1.0$         
            & 50                        
            & y                         
            & 2                         
            & $z_{\rm acc}=10$, $p=1.6$, $u/f=0.014$, plf$=6\times10^{-6}$                          
            & 5.90/5                    
            \\ 
&&&&&&&&&& Av. {\em Chandra} flare, synch. &\\
36 & F
            & $50$        
            & 2.5                         
            & 1.0                       
            & 85                        
            & $5.0$         
            & 0.01                        
            & y                         
            & 2                         
            & PL:$p=2.3$, $\gamma_{\rm e,max}=500$                          
            & 4.512/7                    
            \\ 
&&&&&&&&&& Biggest {\em Chandra} flare, SSC &\\
37 & Q
            & $50$        
            & 2.5                         
            & 1.0                       
            & 85                        
            & $1.3$         
            & 1                        
            & y                         
            & 2                         
            & $z_{\rm acc}=5$, $p=1.2$, $u/f=3\times10^{-7}$,plf$=3\times10^{-3}$                          
            & 14.65/5                    
            \\ 
38 & F
            & $16$        
            & 3                         
            & 1.0                       
            & 85                        
            & $0.6$         
            & 2                        
            & y                         
            & 2                         
            & PL:$p=1.01$,$\gamma_{\rm e,max}=220$                          
            & 4.45/4                    
            \\ 
&&&&&&&&&& $z_{\rm acc}=5$, $p=1.01$, $u/f=3\times10^{-7}$, plf$=5\times10^{-4}$ &\\
&&&&&&&&&& Av. {\em Chandra} flare, SSC &\\
39 & F
            & $16$        
            & 3                         
            & 1.0                       
            & 85                        
            & $0.6$         
            & 2                        
            & y                         
            & 2                         
            & PL: $p=1.01$,$\gamma_{\rm e,max}=220$                          
            & 4.04/7                 
            \\ 
&&&&&&&&&& Av. {\em Chandra} flare, SSC &\\
40 & F
            & $80$        
            & 3                         
            & 1.01                       
            & 85                        
            & $0.6$         
            & 0.1                        
            & y                         
            & 2                         
            & PL: $p=1.5$, $\gamma_{\rm e,max}=500$                          
            & 112/7                    
            \\ 
&&&&&&&&&& Biggest {\em Chandra} flare, SSC &\\
41 & F
            & $25$        
            & 2.5                         
            & 1.01                       
            & 85                        
            & $2.0$         
            & 1                        
            & y                         
            & 2                         
            & $z_{\rm acc}=5$, $p=2.7$, $u/f=3\times10^{-3}$, plf$=0.04$                          
            & 4.193/5                    
\\
&&&&&&&&&& Steep XMM flare, synch. &
\enddata \tablecomments{$^a$ This column describes other adaptations
to the standard model.  ``$\beta$ stretched'' means that we increased
the velocity as a function of distance along the jet by a factor
depending on that distance.  The other comments refer to various ways
of accelerating particles in the jets.  For rows with four additional
parameters, $z_{\rm acc}$ is the location of the acceleration region,
$p$ is the particle index, $u/f$ are plasma parameters which determine
the rate of acceleration, and $plf$ is the fraction of particles
accelerated out of the original quasi-thermal distribution (see
Appendix in \citealt{MarkoffNowakWilms2005} for details).  For rows
with two additional parameters, the particles are assumed to be
accelerated already in the nozzle, in a power law with
$\gamma_{e,min}$ corresponding to the input temperature, particle
index $p$ and maximum lepton Lorentz factor $\gamma_{e,max}$.  \\ $^b$
The $\chi^2$ statistic is sensible for quiescent models only, since it is 
calculated using an averaged quiescent spectrum.  We include the value
for flare models just as a reference.}
\end{deluxetable} 
\twocolumn

\subsection{Analysis Technique}

The jet emission is calculated along its length in increments.  In
order to determine the appearance of the jet on the sky, we calculate
the contribution to the synchrotron spectrum at 43 GHz from each
increment, assumed to be evenly distributed over the radius and
increment width.  Relativistic angle aberration
\citep[e.g.][]{LindBlandford1985} for the increments'
bulk Lorentz factors is taken into account.  This ``profile'' is then
fed into an IDL routine which creates a FITS image of the jet.  Each
model was then rotated by position angles in steps of 15 degrees
covering the full range of angle.  Furthermore, once the jet was
placed with the specified rotation on the image, we convolved it with
a Gaussian ellipse of the scattering as determined below.

\begin{figure}
\centerline{\includegraphics[width=0.5\textwidth]{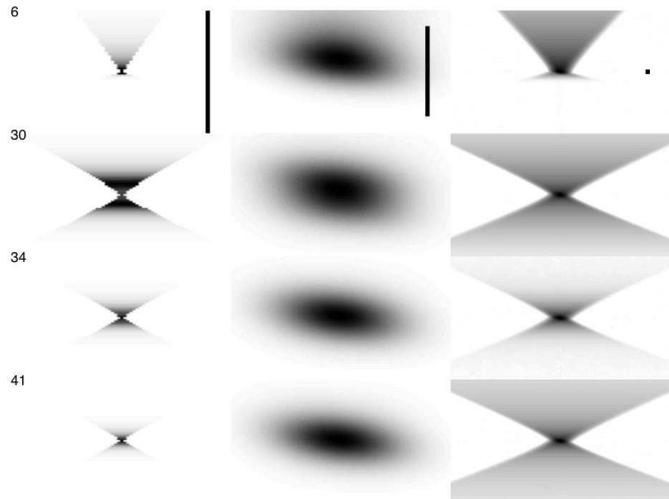}}
\caption{Images of models 6, 30, 34, and 41.  In the left column, we show
the jet model with a linear stretch.  In the middle column, we show the
jet model after it has been convolved with the elliptical Gaussian representing
scattering, also with a linear stretch.  In the  right column, we show the 
convolved jet model in a logarithmic stretch.  The scale bars in the top row
indicate 1 milli-arcsecond.}\label{fig:modeldemo}
\end{figure}

Jet models were imaged on a $2001 \times 2001$ grid with a pixel
resolution of 14 $\mu$arcsec.  Fig.~\ref{fig:modeldemo} shows the
underlying jet model and the resulting scatter-broadened model in
linear and logarithmic scales.  The large-scale differences seen in
the logarithmic representation do not make a significant contribution
to our ability to differentiate between these models, since the total
flux density in the outer regions is very small.  

We directly compare the jet models with high resolution data obtained
at a wavelength of 43 GHz (7 mm).  The data are obtained primarily
with the Very Long Baseline Array (VLBA) and in some cases include a
single Very Large Array (VLA) antenna.  Eight epochs of observations
are described in \citet{Boweretal2004}.  In addition, we include new
observations obtained with the VLBA and the 100m Green Bank Telescope
on 18 May 2004 (experiment code BB183).  These observations were
reduced in the same method as the earlier epochs with calibration for
single-band delay and multi-band delay and rate.

We construct closure amplitude and closure phase from the visibility
data.  The closure phase is the sum of interferometric phases for a triangle
of baselines.  The closure amplitude is a product of interferometric 
amplitudes for baseline quadrilaterals.  Analysis of the closure
quantities is less sensitive than the analysis of calibrated
visibilities because of the reduced number of degrees of freedom.  The
closure quantities are independent of amplitude and phase calibration,
however.  This property which makes them valuable estimators of source
structure that are unbiased by systematic errors in calibration.

In \citet{Boweretal2004} elliptical Gaussian models were fitted to the
closure amplitudes for data sets at wavelengths from 7 mm to 6 cm.
This fitting produced a best-fit elliptical Gaussian as a function of
wavelength.  Combining the VLBI measurements with new measurements of
the size at wavelengths between 17 and 24 cm based on VLA
observations, a size-wavelength relation was determined
\citep{Boweretal2006}.  The scattering ellipse from the long
wavelength observations was computed to be $1.31 \times 0.64$ mas
cm$^{-2}$ in position angle 78$^\circ$.  The size of the ellipse
scales as the wavelength-squared.  Deviations from the
wavelength-squared law at short wavelengths are indicative of the
intrinsic size becoming comparable to the scattering size.  The
magnitude of the scattering ellipse is determined by the spectrum of
turbulent electron density fluctuations.  The orientation and axial
ratio of the scattering ellipse are determined by the magnetic field
properties of the plasma in which the scattering originates.

Each model was directly compared with closure quantities from the data.
First, we added a noise bias to each model image equal to the best-fit value
determined from elliptical Gaussian fitting for the data set.  We
also experimented with using a range of  noise biases that went 
from 0 to 2 times the best-fit value.  We found that the minimum
$\chi^2$ from this procedure was comparable to the $\chi^2$ for the
best-fit noise bias.  Second,
we constructed the image two-dimensional FFT, which is the visibility
plane representation of the data.  Third, 
closure quantities were computed for each model for the time and antenna
sampling of the data set.  Finally, reduced $\chi^2$ was computed for closure 
amplitudes and closure phases for each model and each data set.

In addition to jet models, we also created a model image for an
elliptical Gaussian that represents the best-fit Gaussian from
\citet{Boweretal2004}.  The reduced $\chi^2_\nu=1.9$ from this fit is the baseline
result that jet models must meet or surpass in order to remain
viable.

\begin{figure}
\centerline{\includegraphics[width=0.5\textwidth]{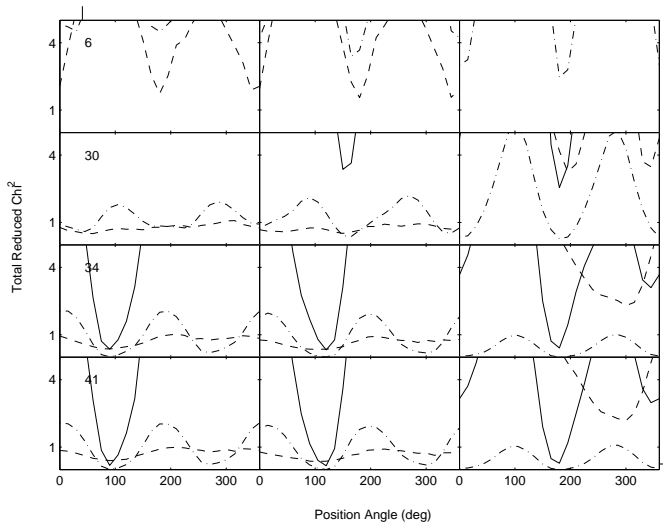}}
\caption{Total $\chi^2_\nu$ as a function of position angle for models 6,
30, 34, and 41 for the case where the data is replaced with closure
quantities calculated from model 41 in position angle 90 deg (left column),
120 deg (middle column), and 180 deg (right column).  Model closure
quantities were computed for three different estimates of the noise,
with the solid line representing the least noise, the dot-dashed line the
middle case, and the dashed line the most noise.}\label{fig:model41}
\end{figure}

To demonstrate the ability of our method to discriminate between
models, we substituted the closure quantities from the data with
closure quantities derived from model 41 in three different position
angles (90, 120, and 180 degrees).  We then compared the substituted
closure quantities with closure quantities from all models and
position angles (Figure~\ref{fig:model41}).  We computed the results
for three different values of the noise bias.  These results show that
we can differentiate between position angles and models in the case of
high signal-to-noise ratios (SNR).

\begin{figure}
\centerline{\includegraphics[width=0.5\textwidth]{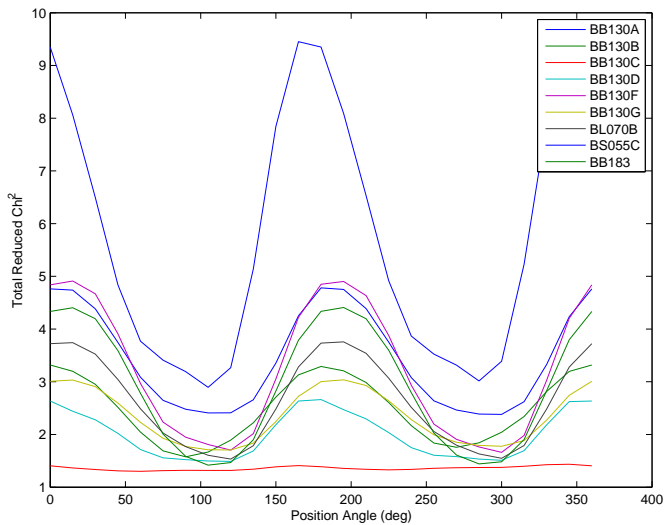}}
\caption{Total $\chi^2_\nu$ as a function of position angle for model 41
showing each radio data set individually.}\label{fig:alldata}
\end{figure}

We also considered whether there are systematic differences in the
model $\chi^2$ between different data sets (Figure~\ref{fig:alldata}).
Seven of nine data sets are essentially consistent with each other.
Data set BB130C shows a flat $\chi^2_\nu$ as a function of position
angle.  This is consistent with larger than average noise
(Figure~\ref{fig:model41}), which was also seen in poor limits from
the Gaussian fitting (Bower et al. 2004).  Data set BS055C shows a
similar profile in $\chi^2_\nu$ versus position angle but
significantly larger values than average.  This suggests that we may
have underestimated the noise for this experiment.  We have 
therefore dropped these two outlier experiments from all further modeling results.

\section{Results}\label{sec5}

\begin{figure}
\centerline{\includegraphics[width=0.5\textwidth]{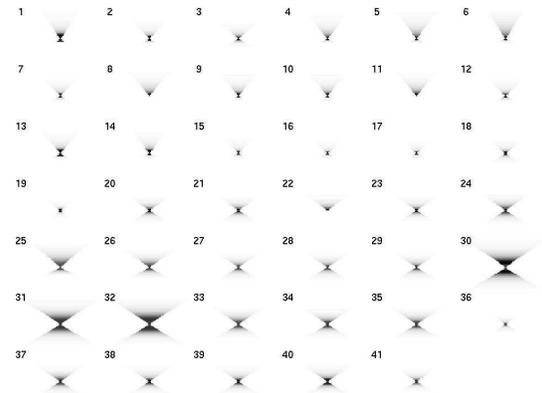}}
\caption{Images of all models prior to convolution with the scattering
ellipse, with a linear stretch.  The scale for each image is 1
milliarcsecond.  }\label{fig:models}
\end{figure}

\begin{figure}
\centerline{\includegraphics[width=0.5\textwidth]{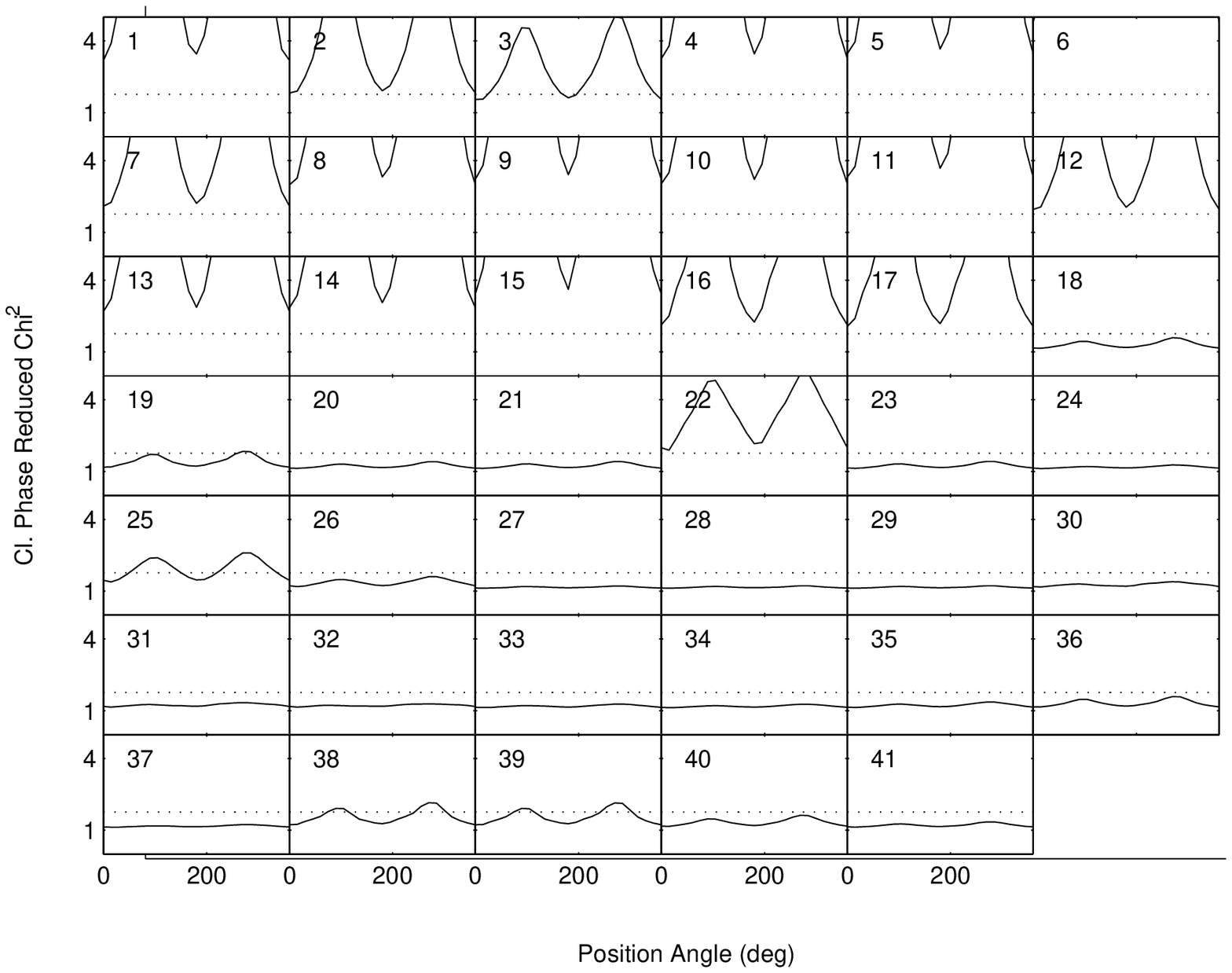}}
\caption{Closure phase $\chi^2_\nu$ as a function of position angle for all models.
The dotted line represents the reduced $\chi^2$ for the best-fit Gaussian
model.
}\label{fig:chi2p}
\end{figure}

\begin{figure}
\centerline{\includegraphics[width=0.5\textwidth]{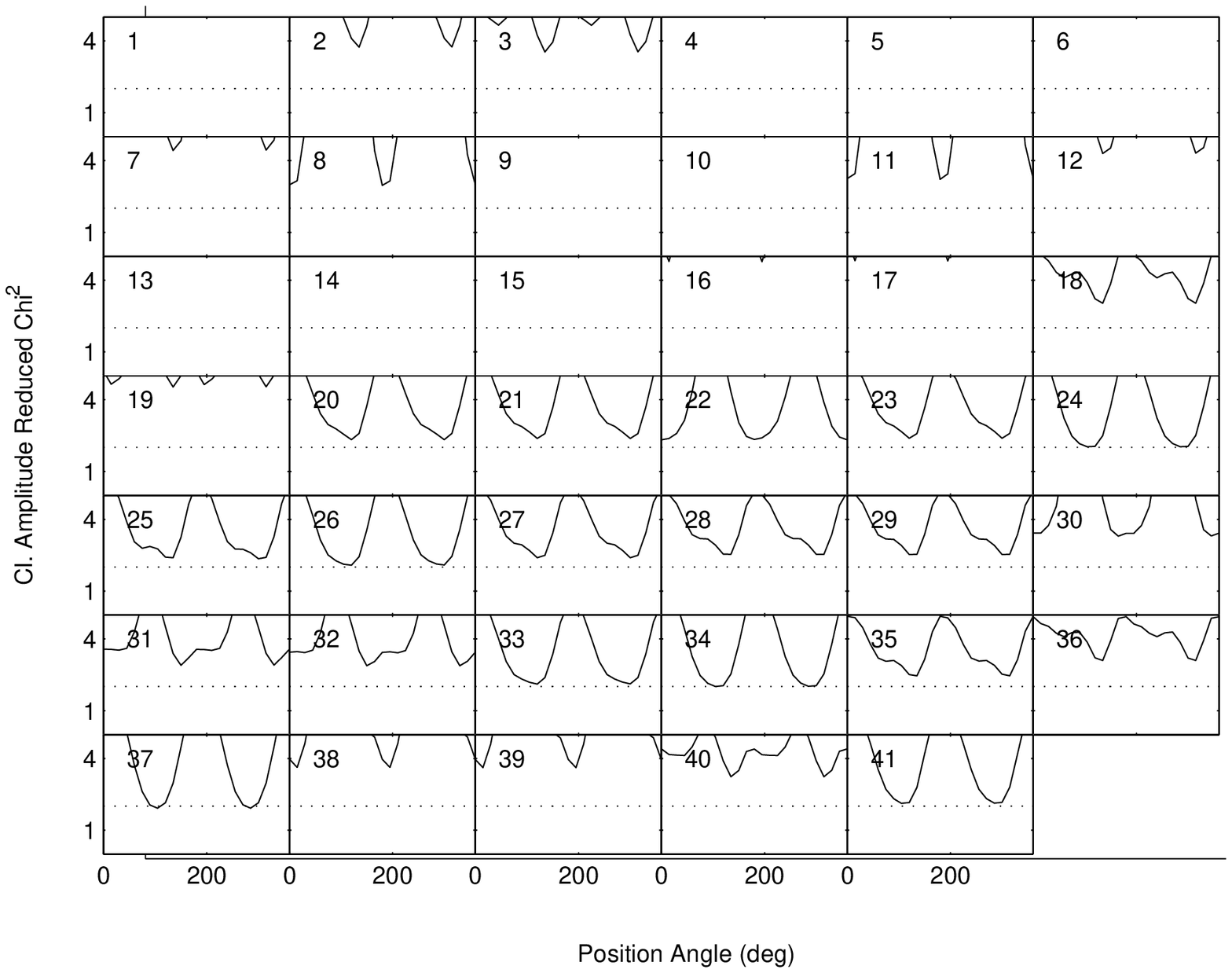}}
\caption{Closure amplitude $\chi^2_\nu$ as a function of position angle for all models.
The dotted line represents the reduced $\chi^2$ for the best-fit Gaussian
model.
}\label{fig:chi2a}
\end{figure}

\begin{figure}
\centerline{\includegraphics[width=0.5\textwidth]{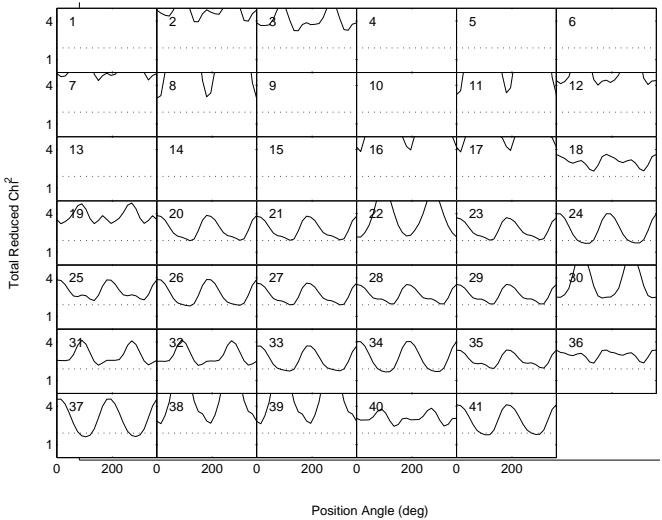}}
\caption{Total $\chi^2_\nu$ as a function of position angle for all models.
The dotted line represents the reduced $\chi^2$ for the best-fit Gaussian
model.
}\label{fig:chi2}
\end{figure}

In Figure~\ref{fig:models} we show all of the model images used in the
analysis.  In Figure~\ref{fig:chi2p}, \ref{fig:chi2a}, and
\ref{fig:chi2} we show the closure phase, closure amplitude, and total
$\chi^2_\nu$ as a function of position angle for each of the models.
In order to see details for the best-fitting models, we plot
$\chi^2_\nu$ only on a scale of 0 to 5.  For several models,
$\chi^2_\nu > 5$; thus where no curve is present, the model is
already strongly rejected.

For a number of models, the minimum $\chi^2_\nu$ is less than or
comparable to the best-fit Gaussian model.  For all cases presented
here, $\chi^2_\nu$ never achieves a significantly smaller value than
the best-fit Gaussian model, which would allow unequivocal rejection
of that model in favor of a jet model.  Instead, these results
demonstrate that we can adequately but not uniquely model the data as
a bipolar, relativistic jet.  This result alone shows that jets in Sgr A*
cannot be ruled out on the basis of their being unresolved.

We see that most of the deviations in $\chi^2_\nu$ as a function of
position angle are visible in the closure amplitude.  For many models,
the closure phase $\chi^2_\nu$ is independent of position angle and is
comparable to the value from Gaussian fitting.  
The closure phase results indicate that any deviations
from axisymmetry in the source image are very small.  The dominant
role of the closure amplitude in variations with position angle is
indicative of sensitivity to the size of the source in a given
direction.

\begin{figure}
\centerline{\includegraphics[width=0.5\textwidth]{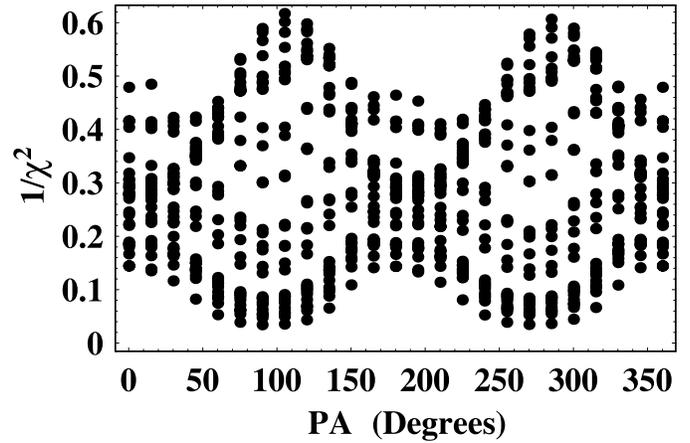}}
\caption{For all models, $1/\chi^2$ as a function of jet PA.  The
  best fit occurs for 105/285$^\circ$. }\label{fig:PAchi2}
\end{figure}

In the East-West direction (90/270$^\circ$), we have the best size
determination because of the better resolution of the telescope in
this direction.  One can think of this as the data having smaller
``error bars'' around these angles.  On the other hand, the scattering
angle is largest, so asymmetries and extended components may be more
obscured.  In the North-South directions, the resolution is worse by a
factor of $\sim 3$.  Thus, minima in $\chi^2$ at 0, 180 and 360 in
several models are not significant.  In these cases the overall fit is
very bad (as can be seen in the regions of better resolution around
90/270$^\circ$) and the dips represent instead regions where our data
are the least constraining.  However in several models we see minima
which are clearly offset from 90/270$^\circ$, such as model 41 where
the minimum occurs at $\sim 105^\circ$ (this can be most easily seen
in Fig.~\ref{fig:PAchi2}).  The peaks seen at 0/180/360$^\circ$
suggest that even with the poorest resolution, the asymmetry is too
great to be consistent with those directions.  The fact that the model
is minimized at an angle where our ability to constrain the asymmetry
is greater is suggestive, and its total $\chi2$ is in fact slightly
lower than the best-fit Gaussian value.  However, it is far too
preliminary to claim a detection of a preferred position angle.  These
results do suggest, however, that with better resolution, especially
in the N-S direction, the position angle of a jet may be constrained,
particularly during flaring states.  Furthermore, many position angles
are clearly ruled out, never achieving close to minimum $\chi^2$ for
any spectrally consistent model.

\begin{figure}
\centerline{\includegraphics[width=0.5\textwidth]{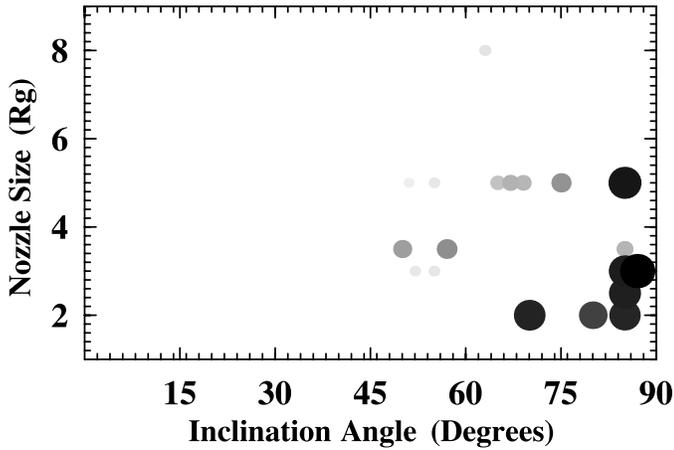}}
\caption{For each of the 30 quiescent models, we plot the minimum
  $\chi^2$ in PA from Fig.~\ref{fig:chi2} for the indicated nozzle
  radius $r_0$ and inclination angle.  The radius and greyscale (from
  white to black) are linear in $1/\chi^2$, and smaller $\chi^2$
  (larger circle) fits are plotted last.  The largest, darkest circles
  have $\chi^2_{\min}\sim1.5$.}\label{fig:scat1}
\end{figure}

\begin{figure}
\centerline{\includegraphics[width=0.5\textwidth]{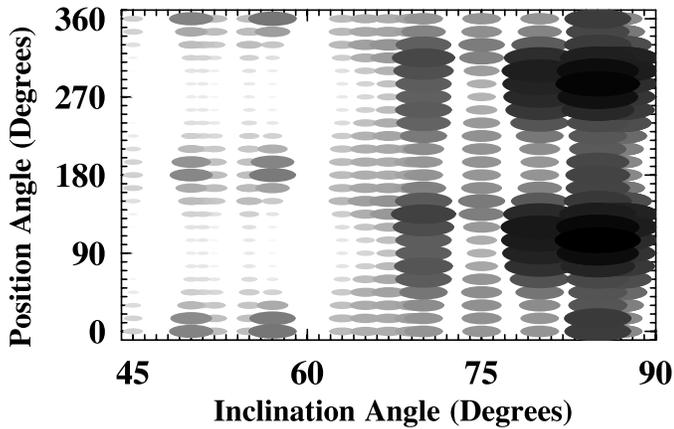}}
\caption{Color scale the same as Fig.~\ref{fig:scat1}, with symbols
  now ellipses (axes scaled linearly in $1/\chi^2$) to better
  illustrate the parameter space, showing $\chi^2$ as a function of model PA and inclination angle.}\label{fig:scat2}
\end{figure}

\begin{figure}
\centerline{\includegraphics[width=0.5\textwidth]{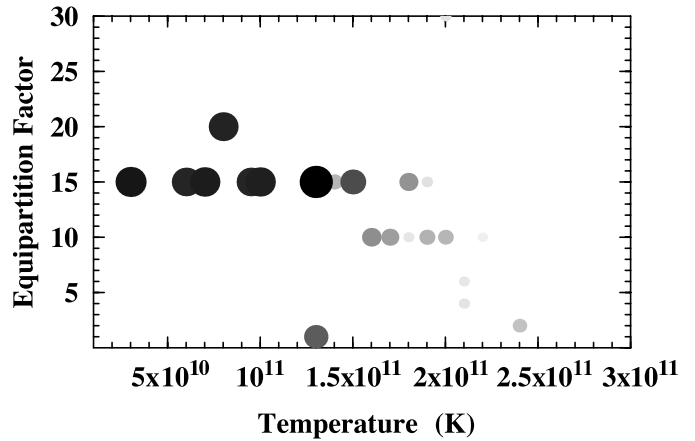}}
\caption{Same symbol definitions as Fig.~\ref{fig:scat1}, showing
  $\chi^2$ as a function of electron temperature and equipartition
  parameter (between magnetic and radiating particle energy densities,
  with $>1$ meaning magnetically dominated).  These are the two most
  important internal rather than geometrical parameters.
  }\label{fig:scat3}
\end{figure}

\begin{figure}
\centerline{\includegraphics[width=0.5\textwidth]{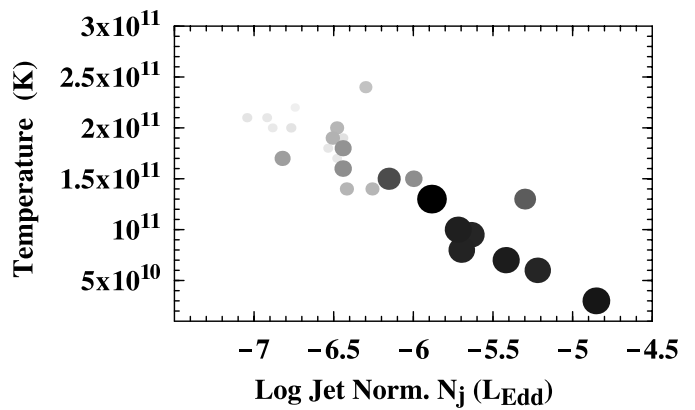}}
\caption{Same symbol definitions as Fig.~\ref{fig:scat1}, showing
  $\chi^2$ as a function of electron temperature and jet normalizing
  power (related to, but slightly less than, the total power; see
  Appendix in \citealt{MarkoffNowakWilms2005}).  There is a clear
  relationship between these two parameters.
  }\label{fig:scat4}
\end{figure}

In Figs.~\ref{fig:scat1}--\ref{fig:scat4}, we show ``scatter plots''
of the minimum $\chi^2$ from Fig.~\ref{fig:chi2} associated with some
model parameters, for the 30 quiescent models only.  The size and
darkness of the circle/ellipse are inversely proportional to the
$\chi^2$, i.e, large and black circles/ellipses are the best fits
while lighter, smaller regions are not.  The two data sets which are
clearly discrepant from the others as discussed above, BB130C and
BS055C, are not included.

Fig.~\ref{fig:scat1} demonstrates the clear selection of compact jets
(whose smallest scale is the nozzle radius $r_0$) and high inclination
angles.  A much more stringent constraint than the spectrum alone is
the combined effect of these two parameters on the jet profile.  While
any compact nozzle less than several $r_g$, or inclination above $\sim
75^\circ$, is acceptable spectrally, the high level of symmetry
required by the VLBI data strongly favors the most compact jets which
are the most beamed out of the line of sight.  Because the jets are
mildly accelerating, the beaming-induced ``dimming'' increases along
the jet axes, thus emphasizing the less elongated nozzle regions.
These results are also a reassuring confirmation because it would be
surprising and somewhat alarming if the jets were so misaligned as to
be pointing significantly towards the Galactic plane in which we
roughly sit.  Fig~\ref{fig:scat2} also compares two geometrical
parameters, this time the position angle on the sky versus the
inclination angle.  The best fit jet is therefore almost perpendicular
to us, with a position angle on the sky of $\sim 105^\circ$.
Fig.~\ref{fig:PAchi2} shows the clear peak in $1/\chi^2$ at this
angle.

Fig.~\ref{fig:scat3} gives an example of how the additional morphology
comparisons can also help constrain internal jet parameters such as
the equipartition of energy and electron temperature.  While the
overall range of spectrally-allowed temperatures spans a decade in
temperature, the upper range clearly does not provide a compact enough
jet profile.  The equipartition parameter however is best constrained
by the spectral fitting, which has already selected a rather narrow
range.  Values $>1$ are magnetically dominated.  

Finally in Fig.~\ref{fig:scat4} we show that some parameter
degeneracies clearly remain despite our new approach.  Here we plot
the electron temperature against the jet normalization parameter
$N_j$.  A clear range of acceptable values exists in both parameters,
demonstrating for instance how a higher temperature electron
distribution can compensate for lower power because of its more
energetic emission.  This can be understood from the critical synchrotron
frequency relationship $\nu_s\propto B \gamma_e^2$, where $B^2 \propto
N_j$ and $\gamma_e \propto T_e$.

\section{Discussion and Conclusions}\label{sec6}

The most important conclusion of this paper is that a jet model, with
reasonable physical assumptions about its geometry and internal
physics, is capable of explaining the radio through IR (and higher,
during flares) spectrum of Sgr A* and {\em not be visible at all to us
as an object with jet-like morphology!}  Aside from the overall low
jet power, the lack of significant particle acceleration implied by
Sgr A*'s IR spectrum would predict extremely compact jet profiles.
Our results demonstrate that the lack of an imaged jet in Sgr A* is
not necessarily due to any absence, but rather stems from a very
compact, weak source combined with the rather extreme scatter
broadening by Galactic electrons, and limits on our spatial
resolution, especially in the N-S direction.
  
However, even without being able to detect a fully elongated
structure, the combination of spectral fitting with constraints from
comparison with VLBI morphology can significantly limit the acceptable
range of parameter space for jet models of Sgr A*.  Figures
\ref{fig:scat1}--\ref{fig:scat4} visually
demonstrate these new limits, which are successful despite the
preliminary exploration of all parameter space.

Not altogether surprisingly, the additional inclusion of size
constraints from VLBI places tighter limits on the model geometry.
For instance, while fitting the quiescent spectrum can only limit the
inclination angle to $\ga 45^\circ$, the addition of VLBI data in
indicates a very clear preference for $\theta_i \sim 90^\circ$.
Similarly, the size constraints from VLBI also narrow the range in jet
nozzle size from $\la 8r_g$ to $\la 5r_g$ with the best fits at the
smaller end.  In the context of jet models, this would require jet
launching to occur very close to the black hole, within the innermost
stable circular orbit (ISCO) for a Schwarzschild black hole.  

Size constraints also indicate that VLBI data can already begin to
constrain the orientation of jets on the sky.  The best fit is found
over a narrow range $90-120^\circ$ centered at $\sim105^\circ$, in a
region where the resolution of the VLBI is good enough to begin
discerning the asymmetry.  The preferred PA is interesting, in that it
could be related to the average position angle of the electromagnetic
fields and thus give further clues about jet geometry.  Recent
observations of variable linear polarization by \citet{Boweretal2005}
and \citet{Marroneetal2006b} observed PA changes of
$30^\circ-60^\circ$ over timescales of days to months.  Infrared
measurements of the polarization during flares also show significant
variability \citep{Trippeetal2007}.  The variability is most likely
intrinsic, although there may be a favored or mean intrinsic
polarization PA in the various wavelengths, though currently they do
not seem to agree with each other.  Confirming both angles may ultimately
provide important information about the helicity of the magnetic field
threading the jets, or near the black hole.

In addition, Figs.~\ref{fig:chi2p}-\ref{fig:chi2} clearly indicate a
dramatic difference in the goodness-of-fit between quiescent and
flaring models for Sgr A*.  This is because the mechanisms involved in
creating the flares
\citep{Markoffetal2001,LiuMelia2002,YuanQuataertNarayan2003} are
either heating or accelerating the radiating particles, which alters
the optical depth and changes the jet profile on the sky.  Our results
strongly argue for further simultaneous X-ray and VLBI (eventually
preferentially in the millimeter regime) monitoring of Sgr A*, where
these methods can strongly limit the contributions of acceleration and
heating, respectively.

In conclusion, we find that the combination of broadband spectral and
morphological constraints gives encouraging and interesting limits on
jet models (or any model) which cannot be obtained by spectral fitting
alone.  In particular, the current difficulty in constraining the
high-energy contribution of the jets because of the dominant quiescent
thermal X-ray emission highlights the need for new approaches.
Including constraints from VLBI images offers a powerful method to
break the current degeneracy in theoretical models for Sgr A*'s
emission, as well as better constraint individual models themselves.

At 43 GHz and below, the key outstanding problem is to measure the two
dimensional structure of Sgr A*.  This requires a careful selection of
North-South baselines that are sensitive to structure on the scale of
a few hundred micro-arcseconds.  However, it is important to note that
electron scattering still acts to symmetrize the data at 43 GHz, thus
mm/submm VLBI could be even more revealing for these types of studies.
The advantage may, however, be offset by the fact that higher
frequencies probe even smaller scales in the jets, which would be
predicted to be as symmetric as an accretion flow.  On the other hand,
mm/submm VLBI will bring us to scales comparable to those probed by
the IR/X-ray flares.  While this will allow us to better observe
simultaneous flares in all three frequency bands, it also raises the
question of how to distinguish the base region of a jet from an
accretion flow.  By identifying structural changes in morphology with
spectral changes in a flare, the approach presented in this paper will
be able to constrain the geometry, particle distributions and emission
mechanisms contributing to the flares.

Finally we emphasize that the results presented here do not include
modifications due to general relativistic effects near the black hole.
For the current resolutions this may not be critical, but as we probe
closer to the innermost regions with higher frequencies, this clearly
needs to be taken into account \citep[e.g.][]{FalckeMeliaAgol2000,BroderickLoeb2005}.

\section*{Acknowledgments}

We would like to thank Michael Wise for help with the scripts for
producing FITs images from the calculations.  H.F. would like to thank
the Miller Institute for hosting his Visiting Miller Professorship at
the Astronomy Department of UC Berkeley, during which significant
progress on this paper was made.


\label{lastpage}
\end{document}